\documentclass{aastex}          
\usepackage{spr-astr-addons}    
\usepackage{xspace}                                

\newcommand{\msun}{\mbox{${\rm M}_{\odot}$}\xspace}
\def\rev#1{{#1}}

\begin{document}
%
\title{Population synthesis of Galactic subdwarf B stars}

\shorttitle{Population synthesis of Galactic sdB stars}
\shortauthors{G. Nelemans}

\author{G. Nelemans\altaffilmark{1}} 

\altaffiltext{1}{ Radboud University Nij\-me\-gen, Department of
  Astrophysics, IMAPP, P.O. box 9010 6500 GL Nij\-me\-gen, the
  Netherlands}

\begin{abstract}
I briefly review the method of population synthesis of binary stars
and discuss the preliminary results of a study of the Galactic
population of subdwarf B stars. In particular I focus on the formation
of (apparently) single sdB stars and their relation to (apparently)
single helium-core white dwarfs. I discuss the merits of mergers of
two helium white dwarfs and interactions with sub-stellar companions
for explaining these single objects. A preliminary conclusion is that
the current observations suggest both mechanisms may contribute, but
that the helium white dwarfs are likely formed in majority from
interactions with sub-stellar companions.
\end{abstract}


%

\section{Introduction}\label{s:intro}

In a volume like this one, no lengthy introduction to subdwarf B (sdB)
stars is needed, as many aspects of their nature are discussed in the
different contributions. Instead I will briefly describe the motivation to
study the Galactic population of sdB stars. Because rather special
circumstances are needed to form an sdB star (the hydrogen envelope
needs to be expelled at the right moment), they offer one of the best
populations to constrain binary-evolution models, in particular the
common-envelope phase \citep[e.g.][]{2003MNRAS.341..669H,hno+07}. 

Secondly, the fact that some sdB stars are pulsators allows
asteroseismological determination of accurate masses and internal
structure, which provides tests of stellar evolution and in particular
an excellent way to constrain their formation \citep[e.g.][and
contributions of Van Grootel and Hu in this
volume]{2008ASPC..392..231F}.

Finally, the relative brightness of sdB stars allows us to study them
to much larger distances than the other \rev{objects that probe the}
late phases of evolution of low- and intermediate-mass stars, the
white dwarfs. Indeed, they are likely responsible for much of the UV
light observed in elliptical galaxies \citep[e.g.][and Han, this
volume]{2007MNRAS.380.1098H}.

However, I think the close link of sdB stars with white dwarfs --
either low-mass helium-core white dwarfs as ``failed'' sdB stars, or
the low-mass C/O core white dwarf descendants of sdB stars -- deserves
more attention. In the remainder of this paper I will discuss the
principle of population synthesis (Sect.~\ref{s:popsynth}), a model
for the Galactic population of sdB stars (Sect.~\ref{s:galpop}) a
discussion of sdB stars in relation to white dwarfs in
Sect.~\ref{s:single} and the probability of finding sdB stars in
binaries with neutron stars or black holes
(Sect.~\ref{s:massive}).

\section{Population synthesis}\label{s:popsynth}

One of the ways to constrain the binary evolution using sdB stars, is
using so called population synthesis \citep[\rev{for a general
description of the different population synthesis codes
see}][]{1983SvA....27..334K,dc87,1988Ap&SS.145....1L,dek90,1991IAUS..143..459Y,kol93,pm94,pv96,1997A&A...317..487V,fbb98,nyp+00,htp02,2003MNRAS.341..669H,2008ApJS..174..223B}. The
basic concept of this method is that a large number of binary
evolution scenarios are calculated using (approximations to) the
evolution of stars \rev{(both single and those affected by binary
interactions)}, in combination with modelling or recipes of binary
interactions. The scenarios, or their relative weighting are chosen in
such a way as to represent assumed initial parameter distributions
(initial primary and secondary mass and orbital period).  The outcome
of this procedure yields a model for the Galactic population of
binaries, that only needs to be normalized in a suitable way to the
estimated total number of (binary) stars in the Galaxy.

\rev{When studying objects using population synthesis one should keep
  in mind} that several crucial ingredients of the models are poorly
  known. In particular the effect of binary interaction and mass
  transfer on the further evolution of the system, the strength of the
  mass loss (in particular for massive stars), the effect of supernova
  explosions, and evolution of peculiar stars that can only be formed
  via binary interactions. In addition, in order to quantitatively
  compare the models to observed populations, the initial parameter
  distributions must be known. The distribution of masses for the most
  massive (primary) component of the binary is usually taken to be the
  initial mass function (assuming the effect of binaries on the
  determination of the IMF is dealt with already). For the secondary
  mass, the observational constraints are of course heavily biased
  towards similar masses, as otherwise the light of the lower-mass
  companion (secondary) is buried by the bright primary. Finally the
  distribution of orbital periods must be patched together from
  spectroscopic and eclipsing binaries for the shorter periods, via
  visual binaries to common proper motion pairs for the wider
  systems. Currently most people use a standard IMF
  \citep[e.g.][]{ktg93}, a flat mass ratio distribution (i.e. for a
  given primary mass, each mass lower has equal probability) and a
  initial semi-major axis distribution that has equal numbers of
  systems in logarithmic intervals from very close binaries to
  typically $\sim 10^4$ AU. The fact that most people use the same
  initial conditions is good for comparing results, but the currently
  used distributions are most likely too simple.

\begin{figure}[tb]
\includegraphics[height=\columnwidth,angle=-90]{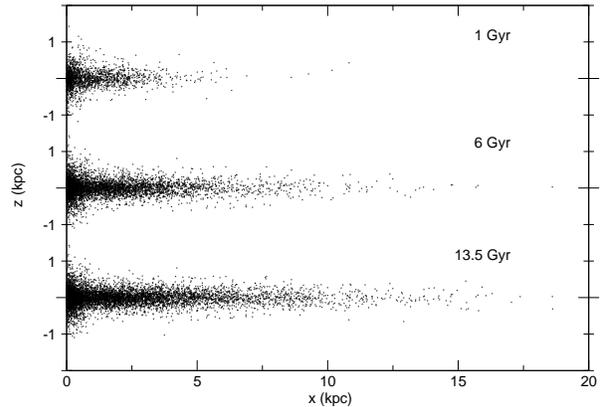}
\caption{Representation of our assumed Galactic star-formation model
  \citep[based on][]{bp99} for three different Galactic ages.  The
  star formation proceeds from the center, including the Bulge, to the
  outer disk. The thick disk and Halo are not modelled.} 
\label{fig:xz}
\end{figure}

When considering low- and intermediate-mass stars also the
normalization is an issue to be considered. Interesting phenomena or
types of systems (such as the formation of sdB stars) may occur
several Gyr after the formation of the main sequence binary. Therefore
the star formation history of the Galaxy becomes important. In most
studies the star formation has been taken to be constant or, for old
populations, to be a single star burst long ago. In \citet{nyp03} we
started to work with a more realistic star-formation history, based on
the galaxy formation models of \citet{bp99}. This is an inside-out
star formation model, that peaked early in the history of the Galaxy
and has decreased substantially since then. In Fig.~\ref{fig:xz} we
show three snap shots of the formation of the Galaxy, including Bulge
and thin disk. The thick disk and Halo are not included in the model.

\section{Galactic populations of sdB stars}\label{s:galpop}

\citet{hpm+02,2003MNRAS.341..669H} and \citet{2005ARep...49..871Y}
have performed a detailed population synthesis of sdB stars, building
on many earlier studies of the formation of helium core burning stars
in binaries \citep[e.g.][]{web84,1990SvA....34...57T}.  So why is it
useful to do another population synthesis study?  Firstly, because of
all the uncertainties in the binary evolution and assumptions about
initial parameters and star formation, it is good to compare similar
calculations from different groups. In particular, we have concluded
from our investigations of the formation of double white dwarfs that a
different description for the common-envelope phase was needed in our
model \citep{nvy+00}, \rev{although this has been used by
\citet{2005ARep...49..871Y} as well}. Secondly, as discussed above, we
use a rather different star formation history.

The model presented here is basically the same as that discussed in
\citet{nyp03}, with the exception that for our alternative common
envelope we use a value of $\gamma = 1.5$, rather than 1.75. This
value fits most of the observed binaries \citep{nt03}. Note that we
still apply the alternative common envelope only to the cases where
none of the two stars is a compact object and the common envelope is
triggered by dynamical unstable Roche-lobe overflow (rather than a
tidal instability). The reason is that, even though in \citet{nt03} we
conclude that the alternative formalism seems to be able to explain
all observed binaries, the motivation for the alternative formalism is
the large amount of angular momentum available in binaries with
similar mass objects.

The basic stellar evolution input is described in
\citet{pv96,nyp+00}. The stellar evolution tracks are somewhat
outdated and we are in the process of updating them. For the moment
the evolution of the core of the star is determined by core-mass
luminosity relations and following of the growth of the core by shell
burning \citep{nyp+00}. \rev{In our population synthesis calculations,
we classify all helium core burning stars that have lost their
hydrogen envelope as sdB stars}.

The only update for this work is that we allow degenerate helium cores
that get exposed \rev{via binary interactions and have a core mass
close to the mass where the helium flash happens} \citep[within
$\sim$0.02 \msun, see][]{hpm+02,hno+07} to ignite and become sdB stars
\citep{ddr+96}. This feature was not included in our models before,
\rev{as those cores were assumed to become helium white
dwarfs}. Because the helium flash \rev{for single stars} in the
stellar models that we use happens at a rather low core mass of 0.446,
\rev{this channels produces sdB stars with masses} between 0.426 and
0.446\msun.

\subsection{Results}\label{s:results}

I present here the preliminary results of a model for the Galactic
population of sdB stars that will be published in more detail
elsewhere. We distinguish three classes of objects, sdB stars with
main sequence (MS) companions, with white dwarf (WD) companions and
single sdB stars.

\begin{table}[tb]
\caption{\rev{Local Space densities and total numbers in the Galaxy}
of different classes of sdB stars and binaries}
\label{tbl:densities}
\begin{tabular}{lrr}
\tableline  
Type & space density & total number \\
&  ($10^{-7}$ pc$^{-3}$) & ($10^5$)\\
\tableline 
sdB + MS \\ ($V_{\rm sdB} \le V_{\rm MS}; P <$10 d) & 1.3 & \\
sdB + MS \\ ($V_{\rm sdB} \le V_{\rm MS}$ all P) & 7.2 &\\
sdB + MS (all) &  23 & 16 \\
sdB + WD \\ ($P <$10 d) & 3.0 & \\
sdB + WD (all $P$) & 5.5 & 6.0\\
He WD mergers & 5.7 & 5.6\\
\tableline 
\end{tabular}
\end{table}

The number of systems in the Galaxy at present is given in local space
densities in Table~\ref{tbl:densities}, \rev{which is
determined by the time and position dependent star-formation rate}. We
give several different values for different selection effects that may
play a role when comparing with observations. First for sdB stars with
MS companions, we select those where the sdB star is at least as
bright as the companion in the V-band. Secondly, for both WD and MS
companions, we give the density of systems with periods shorter than
10 days, which is about the longest period systems that can be found
relatively easily \rev{using spectroscopy} (e.g. Geier, this volume).

The total space density of sdB + MS binaries is about a factor of \rev{5}
larger than that of sdB + WD binaries, but for the ``observable''
systems, the space densities are comparable. The space density of
double helium white dwarf mergers is roughly the same as the sum of
the the ``observable'' binaries. \rev{For comparison with other
population synthesis studies, I show in the last column of
Table~\ref{tbl:densities} the total number of systems in the Galaxy.}

\begin{figure}[tb]
\includegraphics[height=\columnwidth,angle=-90,clip]{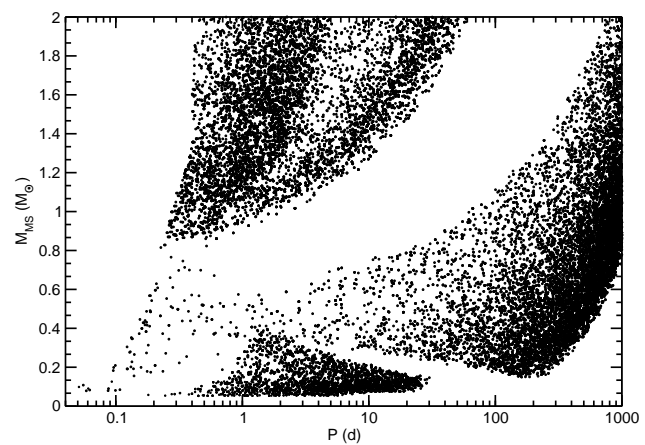}
\caption{Period - main sequence mass distribution for the sdB + MS
  binaries with V-band magnitude \rev{of the binary} brighter than 20
  \rev{but no further selection effects}. The different formation
  channels producing this distribution are discussed in the text.} 
\label{fig:PMms}
\end{figure}

In Fig.~\ref{fig:PMms} I show the distribution of periods and
companion masses for the sdB + MS binaries (without the selection
effects mentioned above, but for a magnitude limited sample). This
gives a nice view on the different formation channels \citep[see
also][]{hpm+02}. The patch at the top, at relatively short periods and
massive companions is a combination of common-envelope ejection of
stars with non-degenerate cores, where relatively large separations
become a lot shorter (darkest region at shortest periods), and stable
Roche-lobe overflow, which in our model can only happen in relatively
short period binaries, when the giant fills its Roche lobe in the
Hertzsprung gap, widening the orbit (wider periods in that
patch). \rev{However, all these systems are ``unobservable'' as the
condition $V_{\rm sdB} \le V_{\rm MS}$ effectively limits the mass of
the MS companion to below $\sim$0.8 \msun}. The patch at the right, at
very long periods, is due to our alternative common-envelope
prescription for stars near the tip of the red giant branch, leading
to mild or hardly any spiral-in. Finally, the triangular patch at the
bottom is for stars that enter a common envelope via tidal
instability, which we treat with the energy balance in our model (see
also Soker, this volume). The low-mass and short period part of that
patch could be associated with the HW Vir type binaries of which there
are several observed with periods around 0.1 d
\citep[see][]{2009ARA&A..47..211H}. The periods on our model are a bit
wider (and less concentrated) than those observed, which may indicate
that we assume a too large efficiency in the common envelope (we
assume $\alpha \lambda = 2$).  A detailed comparison of this model
with the observed systems will given in the forthcoming publication.

\begin{figure}[tb]
\includegraphics[height=\columnwidth,angle=-90]{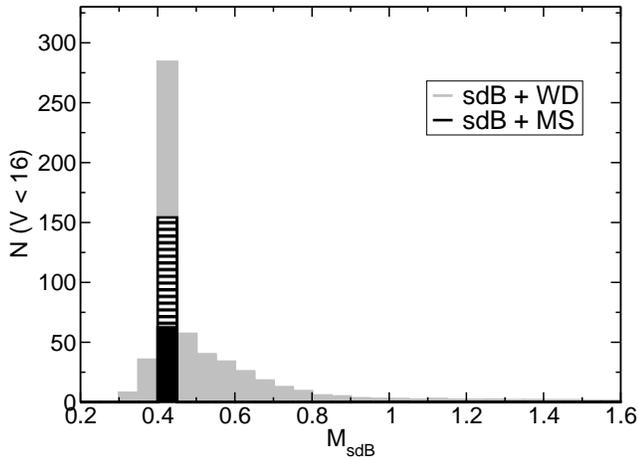}
\caption{Mass distribution of \rev{sdB stars in close binaries. Dark
  histogram (total): sdB + MS stars (with $V_{\rm sdB} \le V_{\rm
  MS}$).  The solid histogram gives the subset of systems with
  low-mass ($\le$0.15 \msun) companions.} Light histogram: sdB + WD
  systems.}
\label{fig:MsdB}
\end{figure}
\begin{figure}[tb]
\includegraphics[height=\columnwidth,angle=-90]{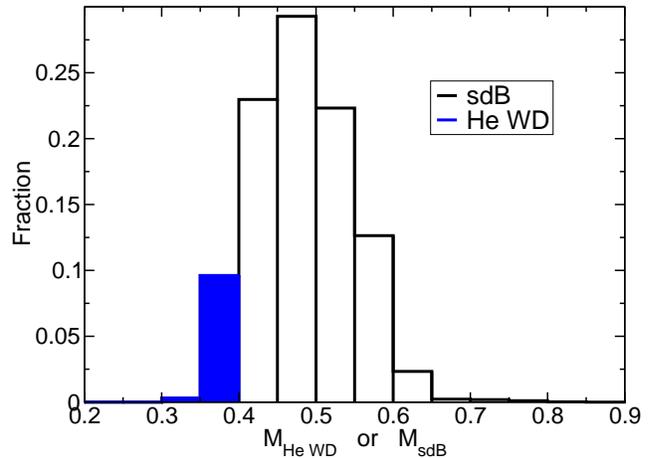}
\caption{Mass distribution of the merger products of double helium
  white dwarfs. \rev{The solid histogram below 0.4 \msun shows those
  merger products that to not have enough total mass to ignite helium
  and become an sdB star. Above 0.4 \msun the ones that do ignite and
  thud become single sdB stars are shown (open histogram)}.}
\label{fig:hdhdmerge}
\end{figure}

\rev{In Figure~\ref{fig:MsdB} I show the mass distribution of close
binary (periods less than 10 days) sdB stars with MS (dark) and WD
(light) companions.}  For the systems with MS companions
I only show systems where the estimated V-band magnitude of the sdB
stars is smaller or equal to that of the MS star.  The dark solid
histogram gives the \rev{subset} of systems with very low-mass
($\le$0.15 \msun) companions, to be compared to the number of HW Vir
stars.

The sdB stars with MS companions all come from \rev{binary
  interactions close to} the tip of the red giant branch and thus have
  \rev{masses between 0.426 and 0.446\msun.}  The ones with WD
  companions can also originate in more massive stars ($M \ga 2\msun$)
  that ignite helium non-degenerately in their cores and thus can both
  be lower as well a higher \rev{mass} than those from the tip of the
  giant branch.  Observable sdB stars with MS companions cannot
  originate from stars with non-degenerate cores \rev{in our models},
  because these stars are typically more compact and, because of their
  larger total mass, have more massive envelopes. \rev{They could}
  only be unbound by \rev{massive} MS companions that would outshine
  the sdB star.  \rev{In Fig.~\ref{fig:hdhdmerge} the distribution of
  single sdB stars that originate from the merger of two helium white
  dwarfs (see Sect.~\ref{s:single})}. The result of mergers are, of
  course, in general more massive.

\section{Single sdB stars and single helium-core white
  dwarfs}\label{s:single}

Here I further discuss the formation of (apparently) single sdB/sdO
stars. Different models have been proposed: enhanced mass loss on the
red giant branch \citep{ddr+96}, the merger of two helium white dwarfs
\citep[e.g.][for sdO stars]{web84} or interactions with low-mass
(sub-stellar) companions \citep{sok98b}. \rev{In the remainder we
discuss the latter two, as it is unclear if the enhanced wind scenario
actually happens.}

Interestingly, in \rev{both} these cases the same mechanism will also
produce helium white dwarfs.  In the case of the merger when the total
mass is too low to ignite the helium \citep[probably around 0.4 \msun,
see][]{hpm+02}, and in the case of an interaction on the giant branch
when the mass of the degenerate core is too low to ignite helium after
the hydrogen envelope is lost \citep[as proposed for the observed
single helium white dwarfs in][]{nt98}. In both cases the ratio of the
formation of single helium white dwarfs and single sdB stars may be
different, and these model ratios can then be compared to the observed
ratio of single helium white dwarfs \rev{to} single sdB stars.

\begin{figure}[tb]
\includegraphics[width=\columnwidth]{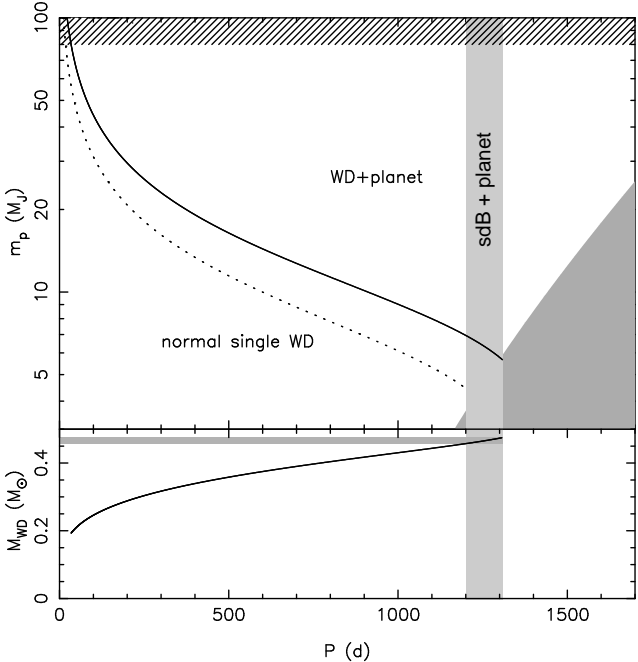}
\caption{Outcome of the interaction of a 1 \msun giant with a
  sub-stellar companion as function of the initial period and mass of
  the companion (top). \rev{Above the solid line the substellar
  companion survives, below it evaporates inside the giant
  envelope. Between the dotted and solid line a significant fraction
  of the envelope mass is lost and the rest may be lost by stellar
  winds. The lower panel gives the mass of the core at the moment of
  the interaction, i.e. the mass of the resulting stars. In order to
  get an sdB star the core needs to be within a very small mass region
  (see Sect.~\ref{s:galpop}), so only systems with initial periods
  between 1200 and 1300 days can produce sdB stars, the rest produces
  a helium white dwarf.  Adapted from \citet{nt98}} }
  \label{fig:planet}
\end{figure}

In Fig.~\ref{fig:hdhdmerge} I show the mass distribution of the
double helium white dwarf merger products which shows that in the
merger case the vast majority ($\sim$90\%) of objects would ignite
helium and thus become sdB (or sdO) stars.  Fig.~\ref{fig:planet},
\rev{adapted} from \citet{nt98}, shows the outcome of a
common-envelope of a sub-stellar companion with a 1 \msun red giant
for different initial orbital periods and companion masses. If I look
at systems in which the sub-stellar companion survives the common
envelope \rev{and the degenerate helium core is massive enough to
ignite and produce an sdB star (denoted sdB+planet in the figure)},
the initial orbital period of the sub-stellar companion is restricted
to a very narrow range, say between 1200 and 1300 days.  Therefore
\rev{if we assume as initial period are equally likely}, the
probability of forming a helium white dwarf is a lot larger than of
forming an sdB star.  A complicating factor is that even if the
sub-stellar companion merges with the giant (region denoted normal
single WD in the figure), it may cause significant mass loss and
spin-up the giant enough that later it will lose its envelope in a
stellar wind, exposing the core either when it is still on the red
giant branch (yielding a helium white dwarf or an sdB star depending
on the mass of the core), or after it contracted to the horizontal
branch following the helium flash \citep[see
also][]{sok98b,2008ApJ...687L..99P}. We therefore included sub-stellar
companions in our population synthesis calculation \rev{by simply
extending the flat mass ratio distribution to include sub-stellar
companions}, and looked at the ratio of the formation of single helium
white dwarfs versus single sdB for interactions on the red giant
branch. \rev{As the mass ratio distribution at these masses and the
fraction of stars with substellar companions are very uncertain, we do
not use this calculation to determine the absolute number of systems,
but only to assess the ratio of helium white dwarf to sdB stars
formation.} As expected, for companions that survive the common
envelope, we find a ratio of about 10 for the number of helium white
dwarfs versus sdB stars, but even for the mergers that subsequently
lose their envelope, we find more helium white dwarfs than sdB stars
(by about a factor 3). The total ratio comes out at 5.5.

\begin{figure}[tb]
\includegraphics[height=\columnwidth,angle=-90,clip]{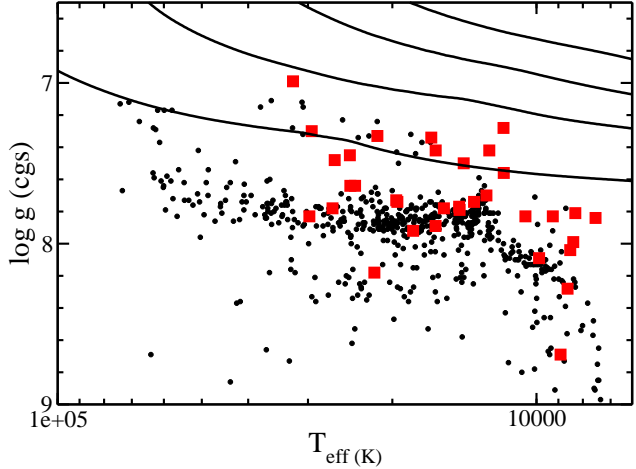}
\caption{Effective temperature -- surface gravity plot for \rev{all}
  white dwarfs observed in the SPY survey \citep{ncd+01}. The squares
  show the close binaries, \rev{the dots the single white dwarfs}. The
  solid lines give colling tracks for \rev{white dwarfs with mass
  0.41, 0.29, 0.24 and 0.20 \msun from \citet{2001MNRAS.323..471A}}}
\label{fig:loggTeff}
\end{figure}

How does this compare with the observational constraints? From the SPY
survey \citep{ncd+01} we find that 15 out of 636 white dwarfs observed
are single helium white dwarfs (see Fig.~\ref{fig:loggTeff}). Assuming
a birth rate of white dwarfs (or planetary nebulae) of $3 \times
10^{-12} {\rm pc}^{-3} {\rm yr}^{-1}$ \citep{pot96}, this results in a
formation rate of single helium white dwarfs of $7.5 \times 10^{-14}
{\rm pc}^{-3} {\rm yr}^{-1}$.  For the single sdB stars we use the
estimated birth rate of $4 \times 10^{-14} {\rm pc}^{-3} {\rm
yr}^{-1}$ \citep{heb86}, which together with a binary fraction of
about 50\% gives a birth rate of $2 \times 10^{-14} {\rm pc}^{-3} {\rm
yr}^{-1}$.  \rev{Because single helium white dwarfs are more common
than single sdB stars, they cannot be predominantly formed from
mergers. Instead it suggests the the single helium white dwarfs are
formed from interaction on the red giant branch.  If all are formed in
that way, the same mechanism will produce single sdB stars at a rate
comparable with the observed rate and thus may be responsible for the
production of most of the single sdB stars.}

However, care should be taken \rev{with this reasoning}, as there are
more ways to produce single sdB stars or single helium white dwarfs
\rev{which could influence the ratio estimated above}. \rev{In
particular, in Sect.~\ref{s:results} I have shown that sdB stars with
low-mass companions in wide binaries can easily be more than half of
the population (Table~\ref{tbl:densities}) even for systems where the
sdB star dominates the (blue) light. Current observational radial
velocity studies are not sensitive to these long periods, i.e. would
not detect the sdB stars as binaries and thus classify them as
single.} In addition, \rev{single sdB stars may be formed in so-called
AM CVn systems \citep[see][for a review]{nel05}. Accretion of helium
onto a helium white dwarf can lead to ignition at the edge and an
outside in helium burning front, very likely turning the accreting
white dwarf into a helium burning star (but probably without H
atmosphere).} A detailed study of this scenario is in
preparation. Finally, \citet{2009A&A...493.1081J} have proposed the
companions of stars that explode as type Ia supernova to be observable
single helium white dwarfs \rev{possibly affecting that side of the
single helium white dwarf to sdB star ratio}. However, the birth rate
derived for single helium white dwarfs above, is about a factor ten
above the rate of type Ia supernova explosions from single degenerate
scenarios, \rev{so this scenario is not important for the overall
number of single helium white dwarfs}.

I conclude that \rev{interaction of sub-stellar companions may be a
significant contributor to the formation of single sdB stars, but} a
full detailed study, including all scenarios and proper selection
effects is needed to come to firm conclusions.

\section{Massive companions?}\label{s:massive}

One of the most intriguing results of the last years is the discovery
of apparently massive, compact companions to sdB stars (e.g. Geier et
al., this volume and references therein). These masses are derived
from the assumption that the sdB stars in these close binaries should
rotate synchronously with the orbit. That means that a combination of
the radial velocity, rotational velocity of the sdB star and an
estimate for the sdB star mass, solves the system and thus gives the
mass of the companion. The worrying aspect is that all systems found
have low inclinations. If the companions are indeed neutron stars or
black holes, as is implied by their inferred mass, their fraction is
astonishingly large, with several objects in a sample of 31 (Geier et
al. this volume). \rev{Indeed \citet{2005ARep...49..871Y} predict
small fractions, and in that study neutron stars and black holes do
not receive an asymmetric kick, which typically drastically reduces
the surviving close binary population.}

Using the same population synthesis model as described before,
including asymmetric kicks at the formation of neutron stars
\citep[see][]{py99,nyp01} we find birth rates for sdB stars with
neutron stars and black holes of $1.6 \times 10^{-5}$ and $2.2 \times
10^{-7} {\rm yr}^{-1}$. This implies about 1 sdB -- neutron star
binary per 100 sdB binaries and 1 sdB -- black hole binary per 10,000
sdB binaries.  It confirms that finding so many massive companions
indeed is not expected, so very exciting if confirmed by higher
inclination systems.

\section{Conclusions}\label{s:concl}

sdB stars are a very useful population for constraining binary
evolution theory via population synthesis studies, because they probe
very particular evolutionary channels and the number of known objects
and their properties is growing rapidly.  I have presented
preliminary results of a population synthesis study of sdB stars and
related objects. Our models show differences with earlier studies,
such as
\citet[e.g.][]{2005ARep...49..871Y,2003MNRAS.341..669H,1990SvA....34...57T}
due to differences in assumptions about binary evolution and the
star-formation history of the Galaxy. This means that when proper
modelling of the selection effects that govern the detectability of
the observed sample is done, we can use the sdB stars to constrain the
binary evolution models.

I have shown the promise of comparing the population of sdB stars
with their direct cousins, such as helium-core white dwarfs and have
compared the birth rates of (apparently) single sdB stars and
(apparently) single helium white dwarfs. I conclude that the birth
rate of single helium white dwarfs is substantially larger than that of
single sdB stars, suggesting that at least for the helium white dwarfs
interactions on the red giant branch with sub-stellar companions are
important.

Finally we have calculated the birth rate of sdB stars with neutron
star and black hole companions and concluded that the models predict
substantially smaller fractions (1 and 0.01 per cent respectively)
than found in short period sdB binaries, assuming the sdB star is in
co-rotation with the orbit. This either means that that assumption is
not correct, or much more interestingly, that there are many mode sdB
stars with massive compact companions in close binaries than the
models predict.

%
\acknowledgments
It is a pleasure to thank Lev Yungelson, Simon Portegies Zwart, Tom
Marsh and Ralf Napiwotzki for much enjoyed collaborations that have
contributed to the material presented here.


%
\bibliographystyle{Spr-mp-nameyear-cnd}  
\bibliography{journals,binaries}                

\begin{thebibliography}{39}
\ifx \bisbn   \undefined \def \bisbn  #1{ISBN #1}\fi
\ifx \binits  \undefined \def \binits#1{#1} \fi
\ifx \bauthor  \undefined \def \bauthor#1{#1} \fi
\ifx \batitle  \undefined \def \batitle#1{#1} \fi
\ifx \bjtitle  \undefined \def \bjtitle#1{#1}\fi
\ifx \bvolume  \undefined \def \bvolume#1{\textbf{#1}}\fi
\ifx \byear  \undefined \def \byear#1{#1} \fi
\ifx \bissue  \undefined \def \bissue#1{#1} \fi
\ifx \bfpage  \undefined \def \bfpage#1{#1} \fi
\ifx \blpage  \undefined \def \blpage #1{#1} \fi
\ifx \burl  \undefined \def \burl#1{\textsf{#1}} \fi
\ifx \doiurl  \undefined \def \doiurl#1{\textsf{#1}} \fi
\ifx \betal  \undefined \def \betal{\textit{et al.}} \fi
\ifx \binstitute  \undefined \def \binstitute#1{#1} \fi
\ifx \bctitle  \undefined \def \bctitle#1{#1} \fi
\ifx \beditor  \undefined \def \beditor#1{#1} \fi
\ifx \bpublisher  \undefined \def \bpublisher#1{#1} \fi
\ifx \bbtitle  \undefined \def \bbtitle#1{#1} \fi
\ifx \bedition  \undefined \def \bedition#1{#1} \fi
\ifx \bseriesno  \undefined \def \bseriesno#1{#1} \fi
\ifx \blocation  \undefined \def \blocation#1{#1} \fi
\ifx \bsertitle  \undefined \def \bsertitle#1{#1} \fi
\ifx \bsnm \undefined \def \bsnm#1{#1} \fi
\ifx \bsuffix \undefined \def \bsuffix#1{#1} \fi
\ifx \bparticle \undefined \def \bparticle#1{#1} \fi
\ifx \barticle \undefined \def \barticle#1{#1} \fi
\ifx \botherref \undefined \def \botherref #1{#1} \fi
\ifx \url \undefined \def \url#1{\textsf{#1}} \fi
\ifx \bchapter \undefined \def \bchapter#1{#1} \fi
\ifx \bbook \undefined \def \bbook#1{#1} \fi
\ifx \bcomment \undefined \def \bcomment#1{#1} \fi
\ifx \oauthor \undefined \def \oauthor#1{#1} \fi
\ifx \citeauthoryear \undefined \def \citeauthoryear#1{#1} \fi
\def \endbibitem {}

\bibitem[\protect\citeauthoryear{{Althaus}, {Serenelli}, and
  {Benvenuto}}{2001}]{2001MNRAS.323..471A}
\begin{barticle}
\bauthor{\bsnm{{Althaus}}, \binits{L.G.}}, \bauthor{\bsnm{{Serenelli}},
  \binits{A.M.}}, \bauthor{\bsnm{{Benvenuto}}, \binits{O.G.}}:
\bjtitle{\mnras}
\bvolume{323},
\bfpage{471}
(\byear{2001})
\end{barticle}
\endbibitem

\bibitem[\protect\citeauthoryear{{Belczynski}
  \textit{et~al.}}{2008}]{2008ApJS..174..223B}
\begin{barticle}
\bauthor{\bsnm{{Belczynski}}, \binits{K.}}, \bauthor{\bsnm{{Kalogera}},
  \binits{V.}}, \bauthor{\bsnm{{Rasio}}, \binits{F.A.}},
  \bauthor{\bsnm{{Taam}}, \binits{R.E.}}, \bauthor{\bsnm{{Zezas}},
  \binits{A.}}, \bauthor{\bsnm{{Bulik}}, \binits{T.}},
  \bauthor{\bsnm{{Maccarone}}, \binits{T.J.}}, \bauthor{\bsnm{{Ivanova}},
  \binits{N.}}:
\bjtitle{\apjs}
\bvolume{174},
\bfpage{223}
(\byear{2008})
\end{barticle}
\endbibitem

\bibitem[\protect\citeauthoryear{{Boissier} and {Prantzos}}{1999}]{bp99}
\begin{barticle}
\bauthor{\bsnm{{Boissier}}, \binits{S.}}, \bauthor{\bsnm{{Prantzos}},
  \binits{N.}}:
\bjtitle{\mnras}
\bvolume{307},
\bfpage{857}
(\byear{1999})
\end{barticle}
\endbibitem

\bibitem[\protect\citeauthoryear{D'Cruz \textit{et~al.}}{1996}]{ddr+96}
\begin{barticle}
\bauthor{\bsnm{D'Cruz}, \binits{N.L.}}, \bauthor{\bsnm{Dorman}, \binits{B.}},
  \bauthor{\bsnm{Rood}, \binits{R.T.}}, \bauthor{\bsnm{O'Connell},
  \binits{R.W.}}:
\bjtitle{ApJ}
\bvolume{466},
\bfpage{359}
(\byear{1996})
\end{barticle}
\endbibitem

\bibitem[\protect\citeauthoryear{de~Kool}{1990}]{dek90}
\begin{barticle}
\bauthor{\bparticle{de~}\bsnm{Kool}, \binits{M.}}:
\bjtitle{\apj}
\bvolume{358},
\bfpage{189}
(\byear{1990})
\end{barticle}
\endbibitem

\bibitem[\protect\citeauthoryear{Dewey and Cordes}{1987}]{dc87}
\begin{barticle}
\bauthor{\bsnm{Dewey}, \binits{R.J.}}, \bauthor{\bsnm{Cordes}, \binits{J.M.}}:
\bjtitle{\apj}
\bvolume{321},
\bfpage{780}
(\byear{1987})
\end{barticle}
\endbibitem

\bibitem[\protect\citeauthoryear{{Fontaine}
  \textit{et~al.}}{2008}]{2008ASPC..392..231F}
\begin{botherref}
\oauthor{\bsnm{{Fontaine}}, \binits{G.}}, \oauthor{\bsnm{{Brassard}},
  \binits{P.}}, \oauthor{\bsnm{{Charpinet}}, \binits{S.}},
  \oauthor{\bsnm{{Green}}, \binits{E.M.}}, \oauthor{\bsnm{{Chayer}},
  \binits{P.}}, \oauthor{\bsnm{{Randall}}, \binits{S.K.}}, \oauthor{\bsnm{{van
  Grootel}}, \binits{V.}}:
In: {U.~Heber, C.~S.~Jeffery, \& R.~Napiwotzki} (ed.)
Hot Subdwarf Stars and Related Objects.
Astronomical Society of the Pacific Conference Series
vol. 392,
p. 231
(2008)
\end{botherref}
\endbibitem

\bibitem[\protect\citeauthoryear{Fryer, Burrows, and Benz}{1998}]{fbb98}
\begin{barticle}
\bauthor{\bsnm{Fryer}, \binits{C.}}, \bauthor{\bsnm{Burrows}, \binits{A.}},
  \bauthor{\bsnm{Benz}, \binits{W.}}:
\bjtitle{\apj}
\bvolume{496},
\bfpage{333}
(\byear{1998})
\end{barticle}
\endbibitem

\bibitem[\protect\citeauthoryear{{Han}, {Podsiadlowski}, and
  {Lynas-Gray}}{2007}]{2007MNRAS.380.1098H}
\begin{barticle}
\bauthor{\bsnm{{Han}}, \binits{Z.}}, \bauthor{\bsnm{{Podsiadlowski}},
  \binits{P.}}, \bauthor{\bsnm{{Lynas-Gray}}, \binits{A.E.}}:
\bjtitle{\mnras}
\bvolume{380},
\bfpage{1098}
(\byear{2007})
\end{barticle}
\endbibitem

\bibitem[\protect\citeauthoryear{{Han} \textit{et~al.}}{2002}]{hpm+02}
\begin{barticle}
\bauthor{\bsnm{{Han}}, \binits{Z.}}, \bauthor{\bsnm{{Podsiadlowski}},
  \binits{P.}}, \bauthor{\bsnm{{Maxted}}, \binits{P.F.L.}},
  \bauthor{\bsnm{{Marsh}}, \binits{T.R.}}, \bauthor{\bsnm{{Ivanova}},
  \binits{N.}}:
\bjtitle{\mnras}
\bvolume{336},
\bfpage{449}
(\byear{2002})
\end{barticle}
\endbibitem

\bibitem[\protect\citeauthoryear{{Han}
  \textit{et~al.}}{2003}]{2003MNRAS.341..669H}
\begin{barticle}
\bauthor{\bsnm{{Han}}, \binits{Z.}}, \bauthor{\bsnm{{Podsiadlowski}},
  \binits{P.}}, \bauthor{\bsnm{{Maxted}}, \binits{P.F.L.}},
  \bauthor{\bsnm{{Marsh}}, \binits{T.R.}}:
\bjtitle{\mnras}
\bvolume{341},
\bfpage{669}
(\byear{2003})
\end{barticle}
\endbibitem

\bibitem[\protect\citeauthoryear{{Heber}}{1986}]{heb86}
\begin{barticle}
\bauthor{\bsnm{{Heber}}, \binits{U.}}:
\bjtitle{\aap}
\bvolume{155},
\bfpage{33}
(\byear{1986})
\end{barticle}
\endbibitem

\bibitem[\protect\citeauthoryear{{Heber}}{2009}]{2009ARA&A..47..211H}
\begin{barticle}
\bauthor{\bsnm{{Heber}}, \binits{U.}}:
\bjtitle{\araa}
\bvolume{47},
\bfpage{211}
(\byear{2009})
\end{barticle}
\endbibitem

\bibitem[\protect\citeauthoryear{{Hu} \textit{et~al.}}{2007}]{hno+07}
\begin{barticle}
\bauthor{\bsnm{{Hu}}, \binits{H.}}, \bauthor{\bsnm{{Nelemans}}, \binits{G.}},
  \bauthor{\bsnm{{{\O}stensen}}, \binits{R.}}, \bauthor{\bsnm{{Aerts}},
  \binits{C.}}, \bauthor{\bsnm{{Vu{\v c}kovi{\'c}}}, \binits{M.}},
  \bauthor{\bsnm{{Groot}}, \binits{P.J.}}:
\bjtitle{\aap}
\bvolume{473},
\bfpage{569}
(\byear{2007})
\end{barticle}
\endbibitem

\bibitem[\protect\citeauthoryear{{Hurley}, {Tout}, and {Pols}}{2002}]{htp02}
\begin{barticle}
\bauthor{\bsnm{{Hurley}}, \binits{J.R.}}, \bauthor{\bsnm{{Tout}},
  \binits{C.A.}}, \bauthor{\bsnm{{Pols}}, \binits{O.R.}}:
\bjtitle{\mnras}
\bvolume{329},
\bfpage{897}
(\byear{2002})
\end{barticle}
\endbibitem

\bibitem[\protect\citeauthoryear{{Justham}
  \textit{et~al.}}{2009}]{2009A&A...493.1081J}
\begin{barticle}
\bauthor{\bsnm{{Justham}}, \binits{S.}}, \bauthor{\bsnm{{Wolf}}, \binits{C.}},
  \bauthor{\bsnm{{Podsiadlowski}}, \binits{P.}}, \bauthor{\bsnm{{Han}},
  \binits{Z.}}:
\bjtitle{\aap}
\bvolume{493},
\bfpage{1081}
(\byear{2009})
\end{barticle}
\endbibitem

\bibitem[\protect\citeauthoryear{Kolb}{1993}]{kol93}
\begin{barticle}
\bauthor{\bsnm{Kolb}, \binits{U.}}:
\bjtitle{\aap}
\bvolume{271},
\bfpage{149}
(\byear{1993})
\end{barticle}
\endbibitem

\bibitem[\protect\citeauthoryear{{Kornilov} and
  {Lipunov}}{1983}]{1983SvA....27..334K}
\begin{barticle}
\bauthor{\bsnm{{Kornilov}}, \binits{V.G.}}, \bauthor{\bsnm{{Lipunov}},
  \binits{V.M.}}:
\bjtitle{Soviet Astronomy}
\bvolume{27},
\bfpage{334}
(\byear{1983})
\end{barticle}
\endbibitem

\bibitem[\protect\citeauthoryear{{Kroupa}, {Tout}, and {Gilmore}}{1993}]{ktg93}
\begin{barticle}
\bauthor{\bsnm{{Kroupa}}, \binits{P.}}, \bauthor{\bsnm{{Tout}}, \binits{C.A.}},
  \bauthor{\bsnm{{Gilmore}}, \binits{G.}}:
\bjtitle{\mnras}
\bvolume{262},
\bfpage{545}
(\byear{1993})
\end{barticle}
\endbibitem

\bibitem[\protect\citeauthoryear{{Lipunov} and
  {Postnov}}{1988}]{1988Ap&SS.145....1L}
\begin{barticle}
\bauthor{\bsnm{{Lipunov}}, \binits{V.M.}}, \bauthor{\bsnm{{Postnov}},
  \binits{K.A.}}:
\bjtitle{\apss}
\bvolume{145},
\bfpage{1}
(\byear{1988})
\end{barticle}
\endbibitem

\bibitem[\protect\citeauthoryear{{Napiwotzki} \textit{et~al.}}{2001}]{ncd+01}
\begin{barticle}
\bauthor{\bsnm{{Napiwotzki}}, \binits{R.}}, \bauthor{\bsnm{{Christlieb}},
  \binits{N.}}, \bauthor{\bsnm{{Drechsel}}, \binits{H.}},
  \bauthor{\bsnm{{Hagen}}, \binits{H.J.}}, \bauthor{\bsnm{{Heber}},
  \binits{U.}}, \bauthor{\bsnm{{Homeier}}, \binits{D.}},
  \bauthor{\bsnm{{Karl}}, \binits{C.}}, \bauthor{\bsnm{{Koester}},
  \binits{D.}}, \bauthor{\bsnm{{Leibundgut}}, \binits{B.}},
  \bauthor{\bsnm{{Marsh}}, \binits{T.R.}}, \bauthor{\bsnm{{Moehler}},
  \binits{S.}}, \bauthor{\bsnm{{Nelemans}}, \binits{G.}},
  \bauthor{\bsnm{{Pauli}}, \binits{E.M.}}, \bauthor{\bsnm{{Reimers}},
  \binits{D.}}, \bauthor{\bsnm{{Renzini}}, \binits{A.}},
  \bauthor{\bsnm{{Yungelson}}, \binits{L.}}:
\bjtitle{Astronomische Nachrichten}
\bvolume{322},
\bfpage{411}
(\byear{2001})
\end{barticle}
\endbibitem

\bibitem[\protect\citeauthoryear{Nelemans}{2005}]{nel05}
\begin{botherref}
\oauthor{\bsnm{Nelemans}, \binits{G.}}:
In: ASP Conf. Ser. 330: The Astrophysics of Cataclysmic Variables and Related
  Objects,
p. 27
(2005).
astro-ph/0409676
\end{botherref}
\endbibitem

\bibitem[\protect\citeauthoryear{Nelemans and Tauris}{1998}]{nt98}
\begin{barticle}
\bauthor{\bsnm{Nelemans}, \binits{G.}}, \bauthor{\bsnm{Tauris}, \binits{T.M.}}:
\bjtitle{A\&A}
\bvolume{335},
\bfpage{85}
(\byear{1998})
\end{barticle}
\endbibitem

\bibitem[\protect\citeauthoryear{{Nelemans} and {Tout}}{2005}]{nt03}
\begin{barticle}
\bauthor{\bsnm{{Nelemans}}, \binits{G.}}, \bauthor{\bsnm{{Tout}},
  \binits{C.A.}}:
\bjtitle{\mnras}
\bvolume{356},
\bfpage{753}
(\byear{2005})
\end{barticle}
\endbibitem

\bibitem[\protect\citeauthoryear{Nelemans, Yungelson, and
  Portegies~Zwart}{2001}]{nyp01}
\begin{barticle}
\bauthor{\bsnm{Nelemans}, \binits{G.}}, \bauthor{\bsnm{Yungelson},
  \binits{L.R.}}, \bauthor{\bsnm{Portegies~Zwart}, \binits{S.F.}}:
\bjtitle{A\&A}
\bvolume{375},
\bfpage{890}
(\byear{2001})
\end{barticle}
\endbibitem

\bibitem[\protect\citeauthoryear{Nelemans, Yungelson, and
  Portegies~Zwart}{2004}]{nyp03}
\begin{barticle}
\bauthor{\bsnm{Nelemans}, \binits{G.}}, \bauthor{\bsnm{Yungelson},
  \binits{L.R.}}, \bauthor{\bsnm{Portegies~Zwart}, \binits{S.F.}}:
\bjtitle{\mnras}
\bvolume{349},
\bfpage{181}
(\byear{2004})
\end{barticle}
\endbibitem

\bibitem[\protect\citeauthoryear{Nelemans \textit{et~al.}}{2000}]{nvy+00}
\begin{barticle}
\bauthor{\bsnm{Nelemans}, \binits{G.}}, \bauthor{\bsnm{Verbunt}, \binits{F.}},
  \bauthor{\bsnm{Yungelson}, \binits{L.R.}}, \bauthor{\bsnm{Portegies~Zwart},
  \binits{S.F.}}:
\bjtitle{A\&A}
\bvolume{360},
\bfpage{1011}
(\byear{2000})
\end{barticle}
\endbibitem

\bibitem[\protect\citeauthoryear{Nelemans \textit{et~al.}}{2001}]{nyp+00}
\begin{barticle}
\bauthor{\bsnm{Nelemans}, \binits{G.}}, \bauthor{\bsnm{Yungelson},
  \binits{L.R.}}, \bauthor{\bsnm{Portegies~Zwart}, \binits{S.F.}},
  \bauthor{\bsnm{Verbunt}, \binits{F.}}:
\bjtitle{A\&A}
\bvolume{365},
\bfpage{491}
(\byear{2001})
\end{barticle}
\endbibitem

\bibitem[\protect\citeauthoryear{{Politano}
  \textit{et~al.}}{2008}]{2008ApJ...687L..99P}
\begin{barticle}
\bauthor{\bsnm{{Politano}}, \binits{M.}}, \bauthor{\bsnm{{Taam}},
  \binits{R.E.}}, \bauthor{\bsnm{{van der Sluys}}, \binits{M.}},
  \bauthor{\bsnm{{Willems}}, \binits{B.}}:
\bjtitle{\apjl}
\bvolume{687},
\bfpage{99}
(\byear{2008})
\end{barticle}
\endbibitem

\bibitem[\protect\citeauthoryear{Pols and Marinus}{1994}]{pm94}
\begin{barticle}
\bauthor{\bsnm{Pols}, \binits{O.R.}}, \bauthor{\bsnm{Marinus}, \binits{M.}}:
\bjtitle{A\&A}
\bvolume{288},
\bfpage{475}
(\byear{1994})
\end{barticle}
\endbibitem

\bibitem[\protect\citeauthoryear{Portegies~Zwart and Verbunt}{1996}]{pv96}
\begin{barticle}
\bauthor{\bsnm{Portegies~Zwart}, \binits{S.F.}}, \bauthor{\bsnm{Verbunt},
  \binits{F.}}:
\bjtitle{A\&A}
\bvolume{309},
\bfpage{179}
(\byear{1996})
\end{barticle}
\endbibitem

\bibitem[\protect\citeauthoryear{Portegies~Zwart and Yungelson}{1999}]{py99}
\begin{barticle}
\bauthor{\bsnm{Portegies~Zwart}, \binits{S.F.}}, \bauthor{\bsnm{Yungelson},
  \binits{L.R.}}:
\bjtitle{\mnras}
\bvolume{309},
\bfpage{26}
(\byear{1999})
\end{barticle}
\endbibitem

\bibitem[\protect\citeauthoryear{Pottasch}{1996}]{pot96}
\begin{barticle}
\bauthor{\bsnm{Pottasch}, \binits{S.R.}}:
\bjtitle{A\&A}
\bvolume{307},
\bfpage{561}
(\byear{1996})
\end{barticle}
\endbibitem

\bibitem[\protect\citeauthoryear{Soker}{1998}]{sok98b}
\begin{barticle}
\bauthor{\bsnm{Soker}, \binits{N.}}:
\bjtitle{AJ}
\bvolume{116},
\bfpage{1308}
(\byear{1998})
\end{barticle}
\endbibitem

\bibitem[\protect\citeauthoryear{{Tutukov} and
  {Yungelson}}{1990}]{1990SvA....34...57T}
\begin{barticle}
\bauthor{\bsnm{{Tutukov}}, \binits{A.V.}}, \bauthor{\bsnm{{Yungelson}},
  \binits{L.R.}}:
\bjtitle{Soviet Astronomy}
\bvolume{34},
\bfpage{57}
(\byear{1990})
\end{barticle}
\endbibitem

\bibitem[\protect\citeauthoryear{{Vanbeveren}, {van Bever}, and {de
  Donder}}{1997}]{1997A&A...317..487V}
\begin{barticle}
\bauthor{\bsnm{{Vanbeveren}}, \binits{D.}}, \bauthor{\bsnm{{van Bever}},
  \binits{J.}}, \bauthor{\bsnm{{de Donder}}, \binits{E.}}:
\bjtitle{\aap}
\bvolume{317},
\bfpage{487}
(\byear{1997})
\end{barticle}
\endbibitem

\bibitem[\protect\citeauthoryear{Webbink}{1984}]{web84}
\begin{barticle}
\bauthor{\bsnm{Webbink}, \binits{R.F.}}:
\bjtitle{ApJ}
\bvolume{277},
\bfpage{355}
(\byear{1984})
\end{barticle}
\endbibitem

\bibitem[\protect\citeauthoryear{{Yungelson} and
  {Tutukov}}{1991}]{1991IAUS..143..459Y}
\begin{botherref}
\oauthor{\bsnm{{Yungelson}}, \binits{L.R.}}, \oauthor{\bsnm{{Tutukov}},
  \binits{A.V.}}:
In: {K.~A.~van der Hucht \& B.~Hidayat} (ed.)
Wolf-Rayet Stars and Interrelations with Other Massive Stars in Galaxies.
IAU Symposium
vol. 143,
p. 459
(1991)
\end{botherref}
\endbibitem

\bibitem[\protect\citeauthoryear{{Yungelson} and
  {Tutukov}}{2005}]{2005ARep...49..871Y}
\begin{barticle}
\bauthor{\bsnm{{Yungelson}}, \binits{L.R.}}, \bauthor{\bsnm{{Tutukov}},
  \binits{A.V.}}:
\bjtitle{Astronomy Reports}
\bvolume{49},
\bfpage{871}
(\byear{2005})
\end{barticle}
\endbibitem

\end{thebibliography}

%

\end{document}